# Life Expectancy at Birth, Estimates and Forecasts in the Netherlands (Females)

## Christos H Skiadas

Technical University of Crete
skiadas@cmsim.net
web: http://www.cmsim.net

**Abstract:** In this paper we explore the life expectancy at birth in the Netherlands by based on a recent theory and a new methodology but also a classical theory of fitting and forecasting. We use the data from 1850 to 2006 provided by the Human Mortality Database (HMD) for the annual deaths per year of age and the structure of the population per year of age. We apply the IM first exit time model which includes also the infant mortality by using the appropriate non-linear regression analysis program that we have developed in Excel. We provide all the related material from the website: http://www.cmsim.net .

## Introduction

Long term predictions of the Life Expectancy and the Life Expectancy at Birth (LEB) is a very critical issue for long range planning for countries and national and international organizations as the World Health Organization (WHO).
Whereas for short term forecasts several tools are in use [7, 10] leading to reliable predictions, the long term estimates are a difficult task. High uncertainty is inherent in exploring the future trends especially in estimating the LEB for various countries. As the LEB is growing for many decades the issue is to explore systematic growth changes leading to better medium and long term forecasts.
An important development is coming from the recent introduction and estimation from demographic data of a deterioration function and the associated maximum deterioration age (Det) (see the recent papers [15, 18, 19] and download from http://www.cmsim.org ). The value of Det is higher than the LEB. However, in nowadays the LEB is growing shortening the gap from Det and in some cases as in Japan (females) the LEB is very close to Det. It is expected that the Life Expectancy at Birth will increase in all the countries thus approaching asymptotically to Det. Another point is the estimate of a future level of the Life Expectancy at Birth by based on the deterioration function which we call the DTR system. The future life expectancy at birth estimated by the DTR system is higher than the Det and provides another level for LEB, whereas Det and DTR approach each other in the long term. Instead to use only the data for the annual values of LEB for making predictions we have two more data sets for Det and DTR thus improving the reliability of forecasts.

## General Theory on Stochastic Modeling of Health State

The first paper related to this theory on stochastic modeling of the health state of an individual was published in 1995 [8]. An application on the Belgium and France data was presented by the authors in the Royal Association of Belgian Actuaries (ARAB/KVBA, founded in 1895) in a meeting in 1995 celebrating the 100 years of the Association.
The modeling approach was focused on finding the distribution of the first exit time of a diffusion process expressing the health state of a person from a barrier. The related theory can be found in [8, 12, 13, 14, 15] and recently in the *International Encyclopedia of Statistical Science*, Springer (2011) [16, 17].
The publications [12, 13, 14, 15, 16] are focused on the development and application of a first exit time model for mortality including the infant mortality. The model termed as First Exit Time-IM is expressed by the following probability density function

$$g(t)=\frac{k(l+(c-1)(bt)^c)}{\sqrt{t^3}}e^{-\frac{(l-(bt)^c)^2}{2t}},$$

where $k$, $l$, $c$ and $b$ are parameters
The parameter $l$ accounts for the infant mortality. A simpler 3 parameter version of this model arises when the infant mortality is limited thus turning the parameter $l$ to be: $l$=0 and the last formula takes the simpler form:

$$g(t)=\frac{k(c-1)(bt)^c}{\sqrt{t^3}}e^{-\frac{(bt)^{2c}}{2t}}$$

In the first model the health state $H(t)$ is expressed by the simple relation

$$H(t)=l-(bt)^c,$$

whereas this formula becomes $H(t)=(bt)^c$ for the second and simpler model.

In both cases a characteristic relation is given by the next function expressing the *curvature* of the health state function during time (time here is the age of the individual)

$$K(t) = \frac{|c(c-1)b^c t^{c-2}|}{(1+c^2 b^{2c} t^{2c-2})^{3/2}}$$

The curvature $K(t)$ gives a measure of the deterioration of the human organism or of loss of *vitality* in terms of Halley [6] and Strehler and Mildvan [20]. We call $K(t)$ the Deterioration Function which provides a bell-shaped curve.

The maximum of the deterioration function (Det) is achieved at the age $T_{Deter}=Deter$ which is given by the formula

$$Det = T_{Deter} = \left[ \frac{(c-2)}{(2c-1)c^2 b^{2c}} \right]^{\frac{1}{2c-2}}$$

A search in several countries shows that Det is slowly changing during the last centuries and practically remaining almost stable until 1950 then showing a steady increase from 1950 until nowadays.

A consequence of the existence of the deterioration function $K(t)$ and its maximum at Det is the postulate that life expectancy at birth will tend to approach Det during time.

The next very important point is to estimate the total effect of the deterioration of a population in the course of the life time termed as *DTR*. This is expressed by the following summation formula:

$$DTR = \int_0^t tK(t)dt \approx \sum_0^t tK(t)$$

Where $t$ is the age and $K(t)$ is the deterioration function.

The last formula expresses the expectation that an individual will survive from the deterioration caused in his organism by the deterioration mechanism. The result is given in years of age and we can construct a Table like the classical life tables. The DTR provides the future trends for the life expectancy and it is very important in doing forecasts along with Det.

The Life Expectancy at Birth (LEB) can also be estimated by a similar summation formula

$$LEB = \int_0^t tg(t)dt \approx \sum_0^t tg(t)$$

The integral gives as an immediate summation whereas the sum in the right hand side of the last formula leads to the estimation of LEB by the classical Life Table method [6, 3, 5, 9].
All the needed estimates are fusible by using a program in Excel which can be downloaded from: http://www.cmsim.net/id13.html . The program estimates the life expectancy from data, from the fitting curve, by calculating the integral and by estimating the future values of the life expectancy by using the DTR method (see the next snapshots from the program; only the first terms are presented here).

**Estimating Life Expectancy from Data**

| Age Category | Census size | Crude Death Rate | Survivorship | Proportional Death Rate | Midpoint Survivorship | Tx=Tx-1-Lx-1 | Life expectancy | Final Life Expectancy |
|---|---|---|---|---|---|---|---|---|
| | | | | | | | | |
| x | nx | dx | lx | qx | Lx=(lx+l(x+1))/2 | Tx=Tx-1-Lx-1 | ex=Tx-lx | ex+0,5 |
| 0 | 1 | 0,00407 | 1 | 0,00407 | 0,9980 | 81,27 | 81,27 | 81,77 |
| 1 | 0,9959 | 0,00038 | 0,9959 | 0,00038 | 0,9957 | 80,27 | 80,60 | 81,10 |
| 2 | 0,9956 | 0,00016 | 0,9956 | 0,00016 | 0,9955 | 79,27 | 79,63 | 80,13 |
| 3 | 0,9954 | 0,00014 | 0,9954 | 0,00014 | 0,9953 | 78,28 | 78,64 | 79,14 |
| 4 | 0,9953 | 0,00011 | 0,9953 | 0,00011 | 0,9952 | 77,28 | 77,65 | 78,15 |
| 5 | 0,9951 | 0,00012 | 0,9951 | 0,00013 | 0,9951 | 76,29 | 76,66 | 77,16 |

**Estimating Life Expectancy from the Fitting Curve**

| Age Category | Census size | Crude Death Rate | Survivorship | Proportional Death Rate | Midpoint Survivorship | Tx=Tx-1-Lx-1 | Life expectancy |
|---|---|---|---|---|---|---|---|
| | | | | | | | |
| x | nx | dx | lx | qx | Lx=(lx+l(x+1))/2 | Tx=Tx-1-Lx-1 | ex=Tx-lx |
| 0 | 1,0000 | 0,003807 | 1 | 0,003807 | 0,9981 | 82,15 | 82,15 |
| 1 | 0,9962 | 0,001346 | 0,9961931 | 0,001351 | 0,9955 | 81,15 | 81,46 |
| 2 | 0,9948 | 0,000733 | 0,9948471 | 0,000736 | 0,9945 | 80,16 | 80,57 |
| 3 | 0,9941 | 0,000476 | 0,9941145 | 0,000479 | 0,9939 | 79,16 | 79,63 |
| 4 | 0,9936 | 0,000341 | 0,9936386 | 0,000343 | 0,9935 | 78,17 | 78,67 |
| 5 | 0,9933 | 0,000259 | 0,9932981 | 0,000261 | 0,9932 | 77,17 | 77,69 |

| Estimating Future Life Expectancy from the Deterioration Curve (DTR Method) | | | | | | | |
|---|---|---|---|---|---|---|---|
| x | K(t) | K(t) Normalised | tK(t) | tK(t) Normalised | Survival Curve | DTR Life Expectancy | qx |
| 0 | 2,671E-15 | 2,687E-15 | 0 | 0 | 1,000000000 | 84,90 | 0,0000000000 |
| 1 | 3,238E-13 | 3,257E-13 | 3,26E-13 | 3,971E-15 | 1,000000000 | 83,90 | 0,0000000000 |
| 2 | 5,359E-12 | 5,391E-12 | 1,08E-11 | 1,314E-13 | 1,000000000 | 82,90 | 0,0000000000 |
| 3 | 3,925E-11 | 3,949E-11 | 1,18E-10 | 1,444E-12 | 1,000000000 | 81,90 | 0,0000000000 |
| 4 | 1,839E-10 | 1,850E-10 | 7,40E-10 | 9,022E-12 | 1,000000000 | 80,90 | 0,0000000000 |
| 5 | 6,496E-10 | 6,536E-10 | 3,27E-09 | 3,984E-11 | 1,000000000 | 79,90 | 0,0000000000 |
| 6 | 1,888E-09 | 1,900E-09 | 1,14E-08 | 1,389E-10 | 1,000000000 | 78,90 | 0,0000000001 |
| 7 | 4,758E-09 | 4,787E-09 | 3,35E-08 | 4,085E-10 | 1,000000000 | 77,90 | 0,0000000004 |
| 8 | 1,075E-08 | 1,082E-08 | 8,65E-08 | 1,055E-09 | 0,999999999 | 76,90 | 0,0000000011 |
| 9 | 2,229E-08 | 2,243E-08 | 2,02E-07 | 2,461E-09 | 0,999999998 | 75,90 | 0,0000000025 |
| 10 | 4,312E-08 | 4,338E-08 | 4,34E-07 | 5,288E-09 | 0,999999996 | 74,90 | 0,0000000053 |

The next two Figures (1 and 2) illustrate special cases for Netherlands (1930 and 2006). The data points, the fitting curve $g(t)$ and the deterioration function appear.

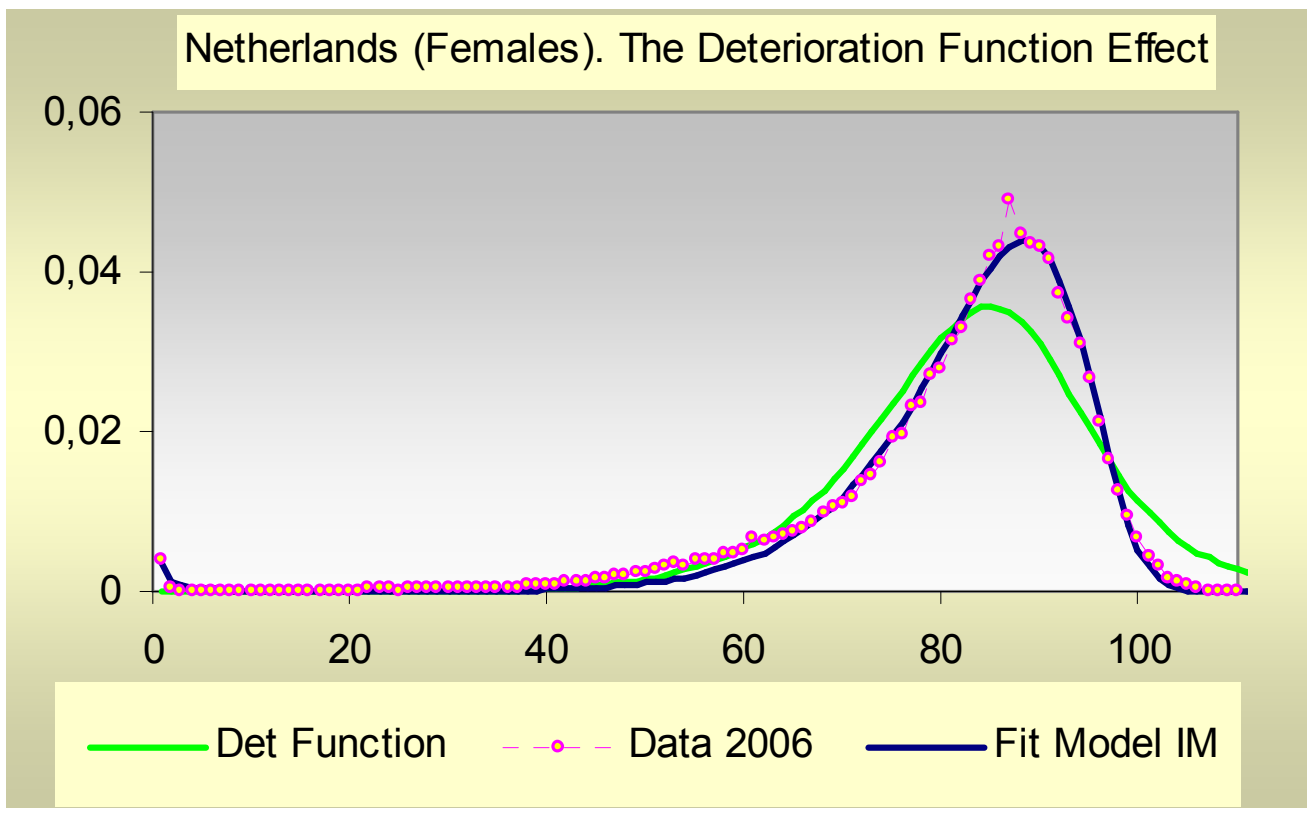

Figure1. Application in Netherlands for 2006

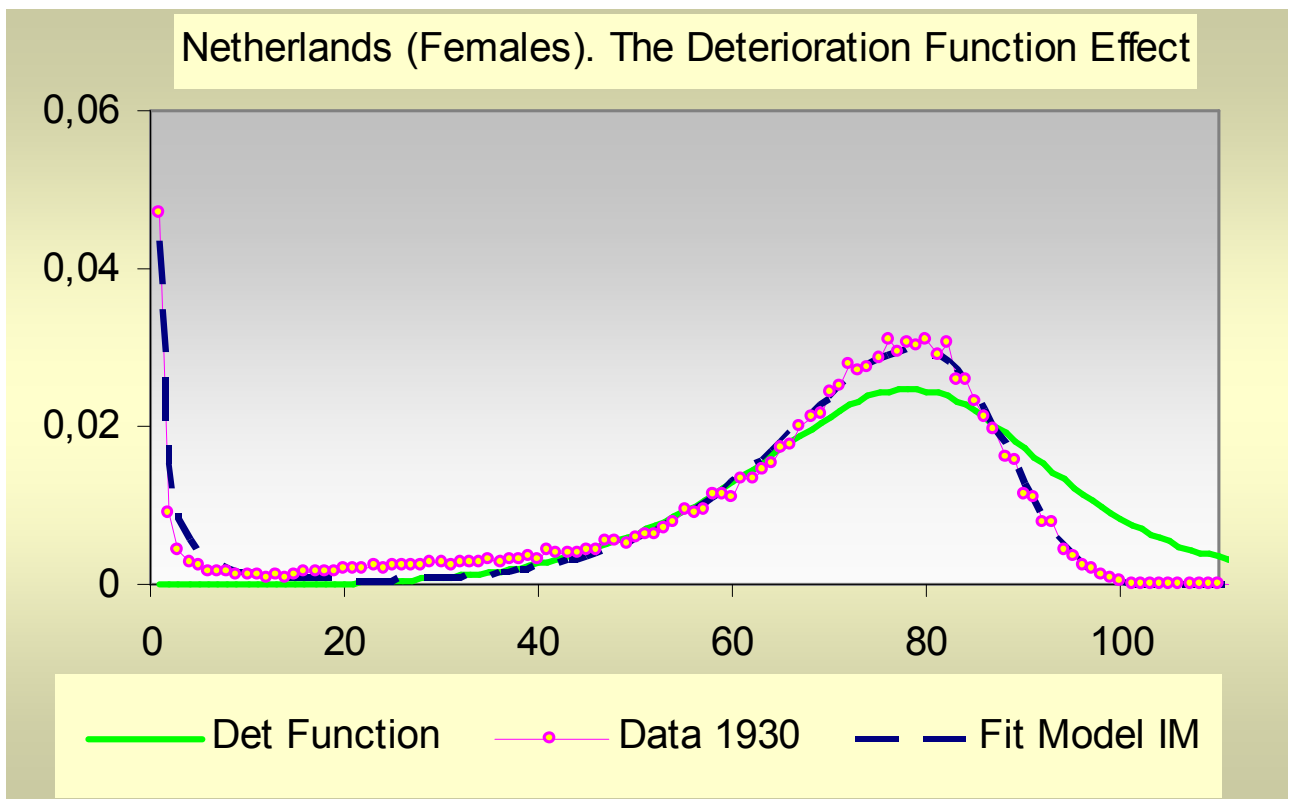

Figure2. Application in Netherlands for 1930

## Modeling and Applications in the Netherlands (Females) First Method of Forecasts

In the following we do forecasts for the Life Expectancy at Birth by using the data provided by the Human Mortality Database (HMD). The population and death data (per age) for females are introduced into the IM-First Exit Time [12, 13, 14] nonlinear regression analysis program (http://www.cmsim.net/id20.html ) estimating all the necessary parameters including LEB (from data, from fitting and the mean value), Det and DTR. The results are summarized in the related Table X at the end of this document.

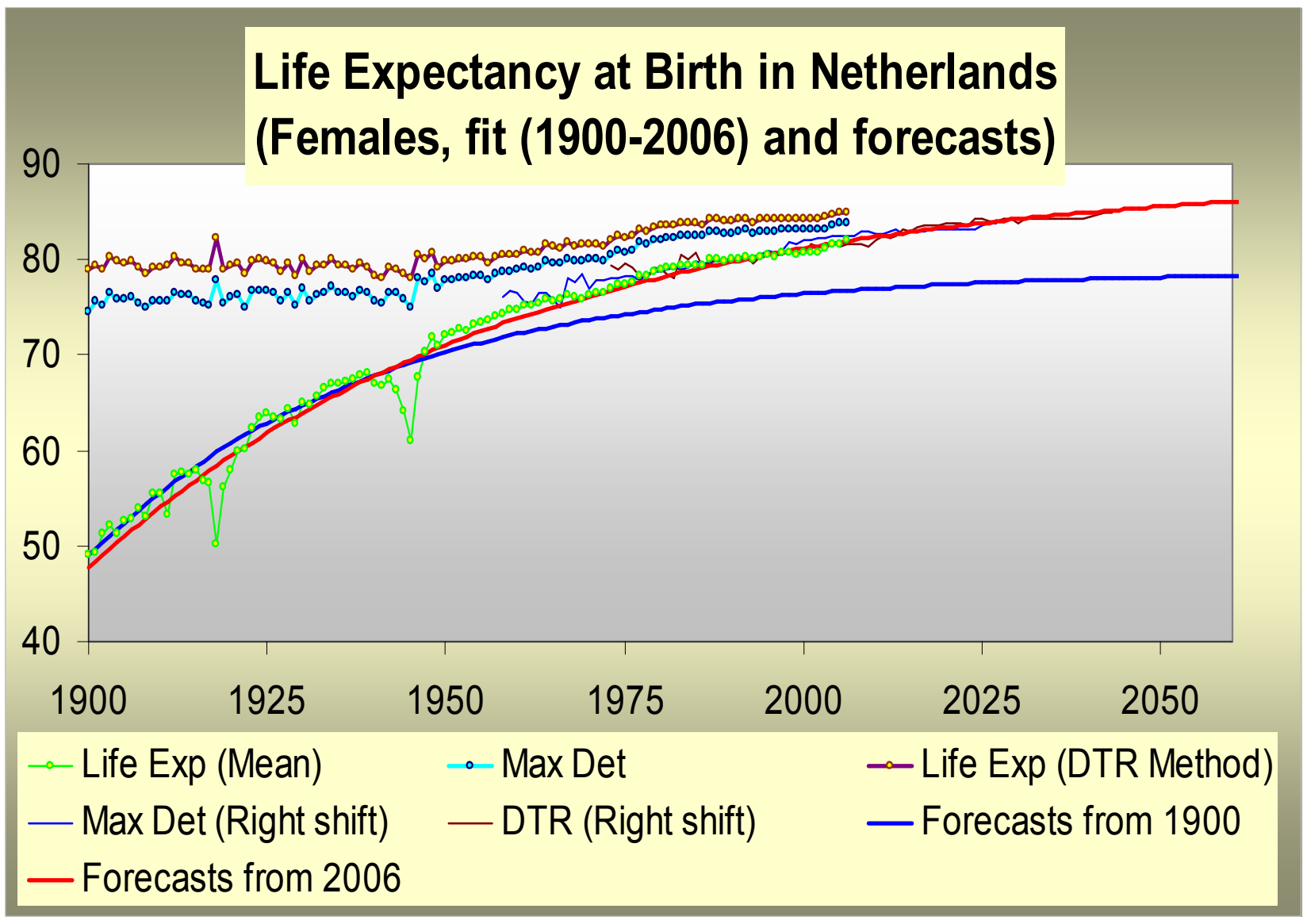

Figure 3. Fit and Forecasts

We use the population and death data from the Human Mortality Database for Netherlands (females) from 1900 to 2006.
We estimate the Life Expectancy at Birth (LEB), the Maximum deterioration age (Det) and the future life expectancy based on the DTR Method (DTR). As it can be easily verified from applications in several countries the LEB tends to increase during time approaching the Det which tends to coincide with the DTR as it is illustrated in Figure 4 where the differences (Det-LEB), (DTR-LEB) and (DTR-Det) appear. By observing Figure 4 it is clear that a systematic relation between the three estimates is present.

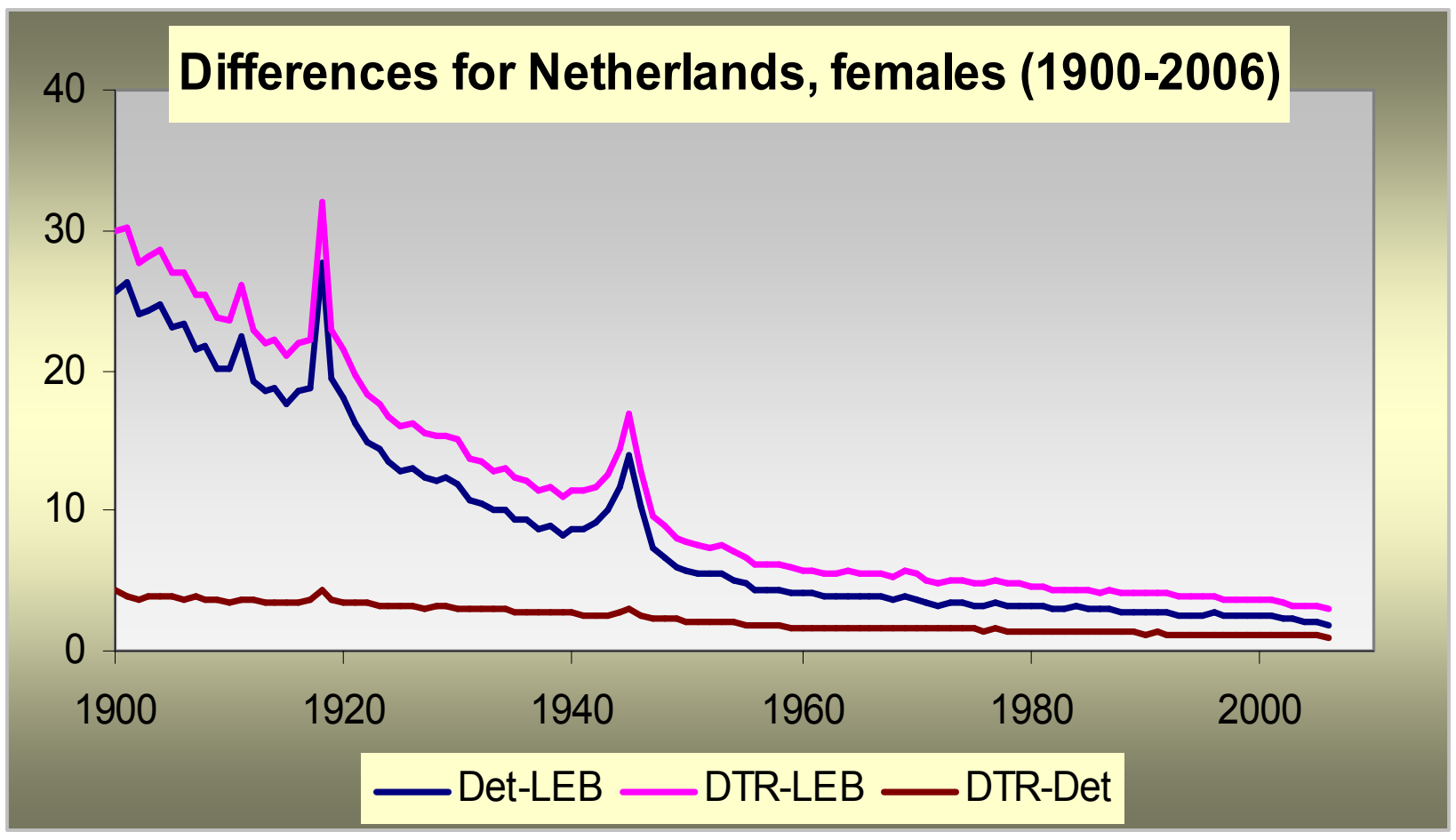

Figure 4. Differences between LEB, Det and DTR

In the following we will use the Det and DTR estimates along with the LEB to improve life expectancy predictions.

The key point is the deterioration function and the maximum point of this function corresponding to the age with the maximum deterioration of the human organism (see the main ideas in [1, 2, 4, 6, 8, 14, 18, 19, 20]). As far as the other causes of death will be diminished due to the improvements in science and the social and community efforts to improve our way of living, the LEB age will by close to the Det age, whereas the future life expectancy estimated by the DTR system will approach the Det values. In other words the future values of the life expectancy at birth can be found by simply shifting the Det and DTR trends to the right as it is illustrated in Figure 3. We can immediately have an estimate for the future trend of LEB by starting from the value of LEB for 1900 ($f_{0=}$48,98 years) and assuming an upper limit F(1900)=DTR(1900)=78,86. The next step is to find the parameter *b* of the next equation as to fit the first data sets for LEB after 1900. The first 10 data points (1900-1909) are sufficient for a good prediction of LEB for almost 40 years (see Figure 3, blue line).

The next step is to find the time lag between LEB and the right shift of the Det. After shifting the Det to the right (see Figure 3) we shift the DTR to the right as to fit on the “right shifted Det”. We thus form a series (by averaging) composed from the values of LEB and the shifted Det and DTR. Then we use this series as an entry in a nonlinear regression analysis program for the estimate of the best fit. As the system tends to stabilize to an upper limit we use the following equation:

$$LEB(t) = F - (F - f_o)\, e^{-at}$$

Where $F$ is the upper limit, $f_0$ is the starting point and $a$ is a parameter expressing the speed of growth. The regression analysis gave the following values

TABLE I

| $F$ | $f_0$ | $a$ |
|---|---|---|
| 88,89 | 47,77 | 0,01676 |

The LEB is estimated from 1900-2006 data for females in the Netherlands. The estimates based on the 1950-2006 data for females in Netherlands are also estimated. For the later case the estimates are also illustrated in Figure 5. The LEB, Det and DTR values are estimated from 1950-2006. Det and DTR are shifted to the right as in the previous case and estimates are done from 2006. The fitting by using the previous equation and the nonlinear regression gave the following values (Table II). The upper limit is lower (87,70 years) than in the previous case (88,89 years).

TABLE II

| $F$ | $f_0$ | $a$ |
|---|---|---|
| 87,70 | 71,75 | 0,01862 |

The predicted values by using the data from 1900 to 2006 and 2050 to 2006 are included in Table X in the end of the paper. Characteristic values are included in the next Table III

TABLE III

| Year | Life Expectancy at Birth, Estimates (Data from 1900-2006) | Life Expectancy at Birth, Estimates (Data from 1950-2006) |
|---|---|---|
| 2010 | 82,37 | 82,48 |
| 2015 | 82,90 | 82,95 |
| 2020 | 83,38 | 83,37 |
| 2025 | 83,82 | 83,75 |
| 2030 | 84,23 | 84,10 |
| 2035 | 84,60 | 84,42 |
| 2040 | 84,95 | 84,71 |
| 2045 | 85,27 | 84,98 |
| 2050 | 85,56 | 85,22 |
| 2055 | 85,82 | 85,44 |
| 2060 | 86,07 | 85,64 |

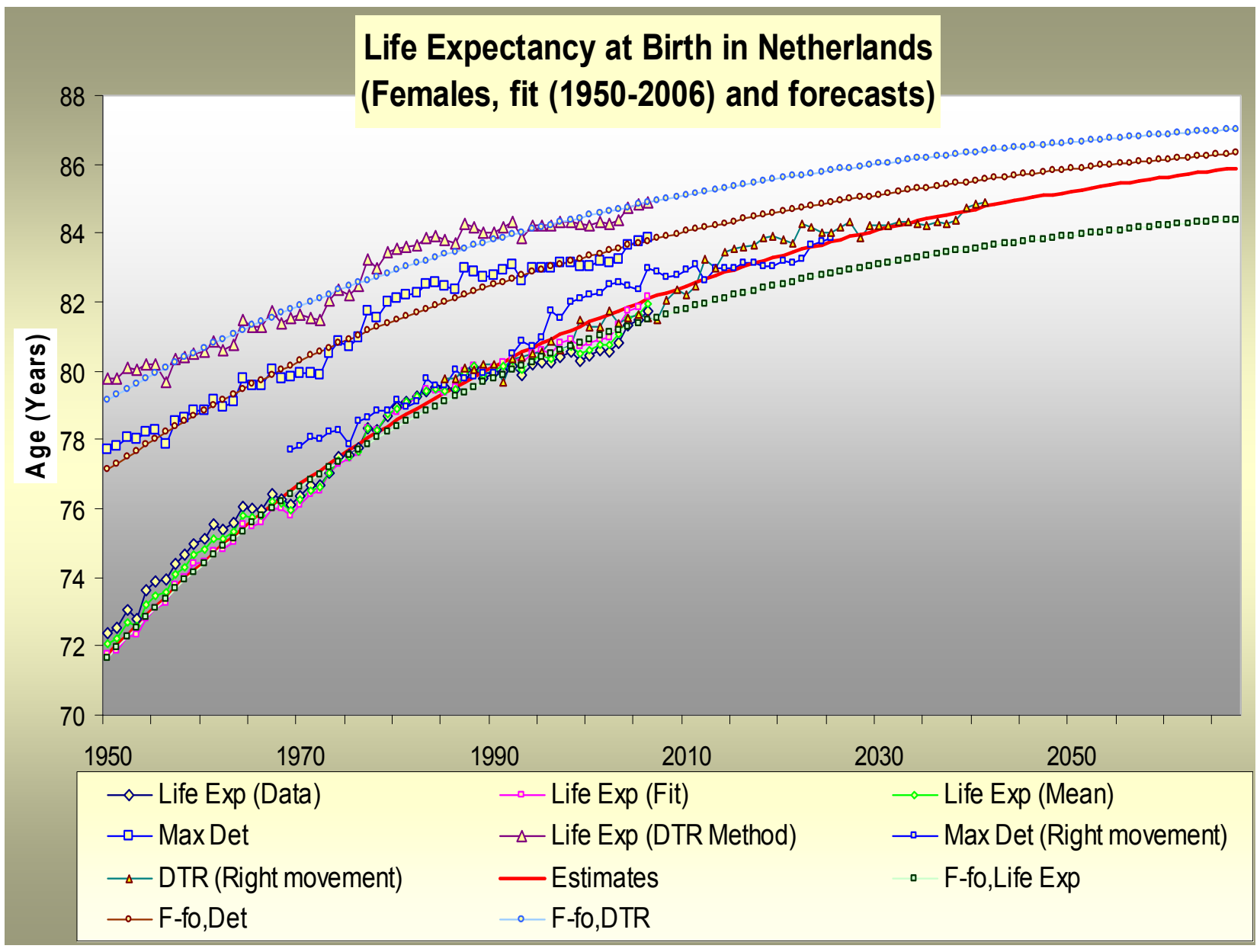

Figure 5. Fit and Forecasts from 2006 in Netherlands

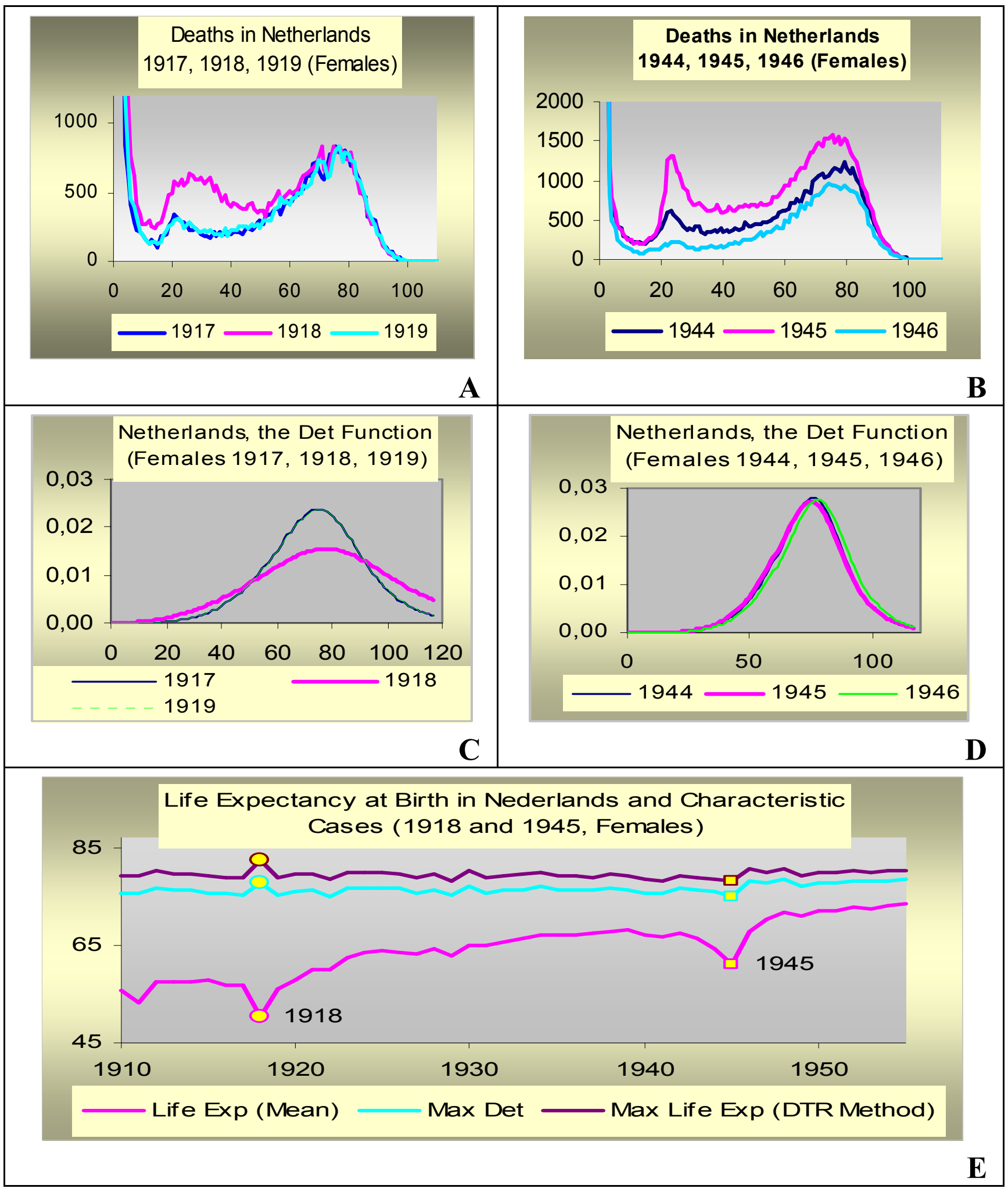

Figure 6 (A, B, C, D, E). Special cases

We turn out to explore two special cases of very low life expectancy at birth in Netherlands that of the year 1918 due to the influenza pandemic and of 1945 due to the after the Second World War effects. The resulting values are compared to the Det and DTR values for the same time period. As it is presented in Figure 6E both Det and DTR show almost stable behavior as they represent the effects of deterioration of the human organism. The year 1945 a mortality excess appear (see Figure 6B) distributed to all the ages. It is demonstrated by a local minimum in the LEB diagram. The corresponding influence to Det and DTR is limited as it was expected by the related theory. More interesting is the case of the influenza pandemic in 1918. As it is illustrated in Figure 6A the mortality excess is distributed in the age group from 15 to 50 years approximately. The resulting value for LEB is a local minimum as it was expected, whereas, Det and DTR show a local maximum. This can be explained by observing Figure 6C. The graphs expressing the deterioration function for the years 1917 and 1919 almost coincide whereas the related graph for the year 1918 is completely different expressing the influence of pandemic influenza. Instead for the years 1944, 1945 and 1946 the three curves for the deterioration function are very close each other (Figure 6D).

## Parameter Analysis of the IM-Model
## Second Method of Forecasts (Classical)

By using the $g(t)$ formula for the death distribution we apply the nonlinear regression analysis program to the Netherlands data for females from 1850 to 2006 and we estimate the parameters of the model. The parameters show systematic changes over time so that future predictions are possible. The parameter $b$ follows an exponential decay process presented in Figure 7. The fitting parameters are summarized in Table IV. The parameter $b$ approaches the lower limit at $b$=0,01175.
The last period (1960-2006) can also be modeled by a simple line of the form $b=a+ct$ where $a$=0,01594 and $c$=-0,00003116. The linear form can apply for short and medium term predictions.

TABLE IV

| $F$ | $f_0$ | $a$ |
|---|---|---|
| 0,01175 | 0,02509 | 0,010074 |

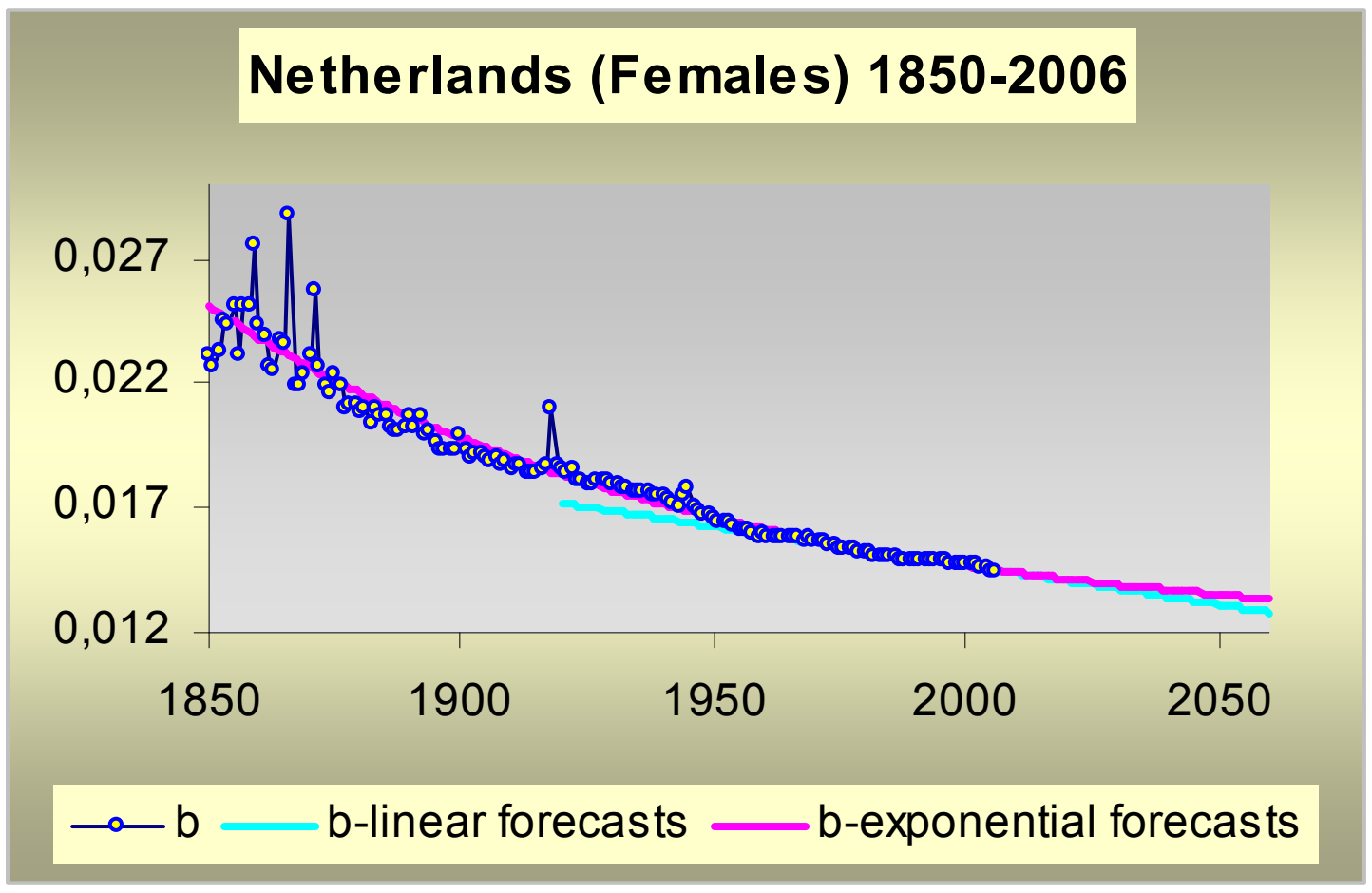

Figure 7. Parameter $b$: fit and forecasts

The parameter $l$ accounts for the infant mortality. As it is illustrated in Figure 8 this parameter follows a negative exponential process tending to very low values close to zero. The nonlinear regression analysis fitting gave the next values for the negative exponential function applied (Table V). The limit of the parameter $l$ is found to be $l$=0,001329 indicating the successful application of the social health services in the infant section.

TABLE V

| $F$ | $f_0$ | $a$ |
|---|---|---|
| 0,001329 | 0,2505 | 0,04329 |

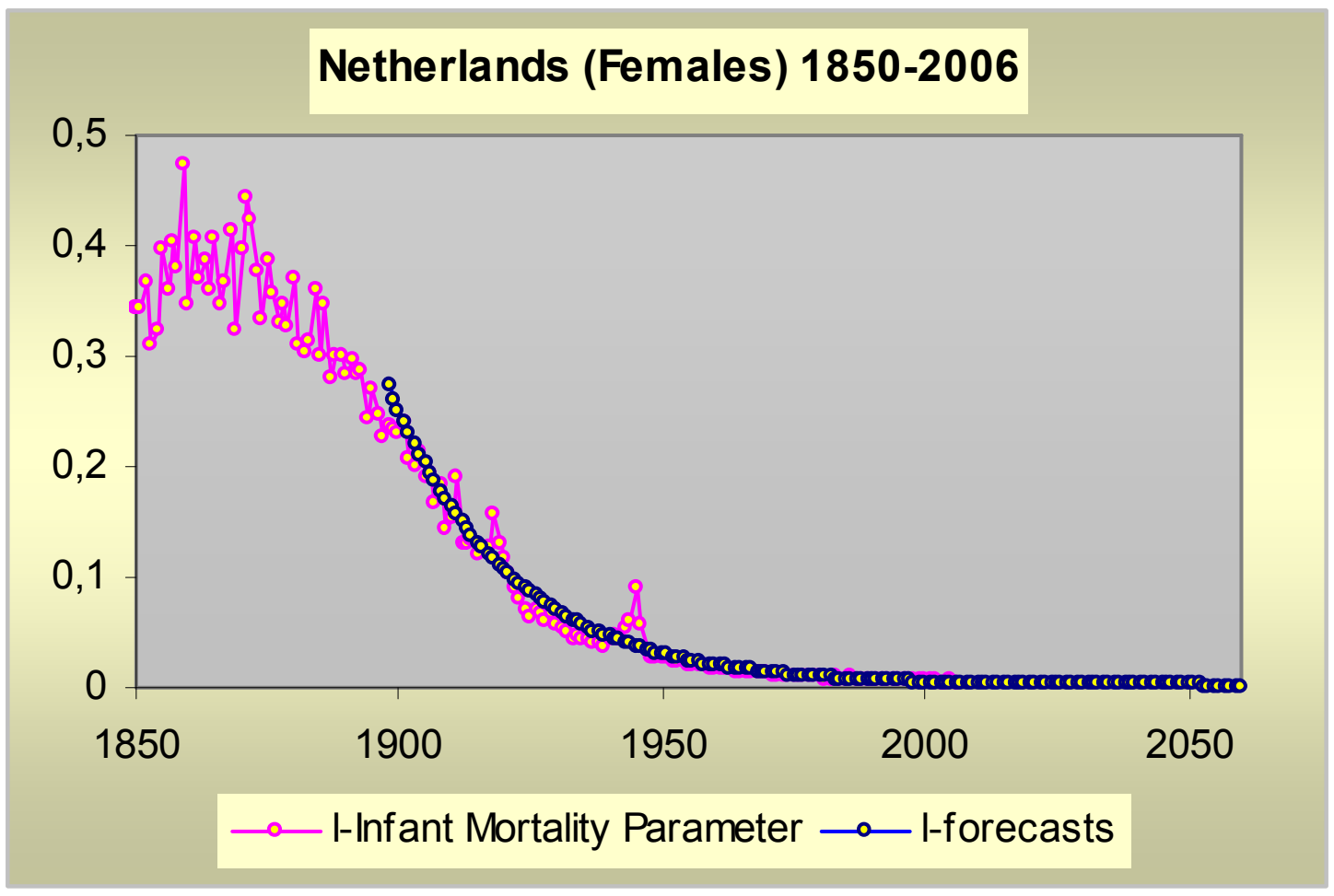

Figure 8. Infant Mortality parameter $l$

The parameter $c$ follows a growing process as it is illustrated in Figure 9. The total period explored is between 1850 and 2006. This period is divided in two periods (1850-1963) and (1964-2006) by observing the related data (see Figure 9). The following exponential function is applied for the two periods

$$Y = F + (f_o - F)e^{at}$$

The parameters from the nonlinear regression analysis fitting are summarized in Table VI. The forecast based on the period (1850-1963) suggests higher values for $c$ than the forecast based on the (1964-2006) data. However, both curves cross each other at 2088 (see Figure 9).

TABLE VI

| Time Period | *F* | $f_0$ | *a* |
|---|---|---|---|
| 1850 -1963 | -1,2738 | 3,7108 | 0,005235 |
| 1964-2006 | 6,9492 | 7,7386 | 0,01969 |

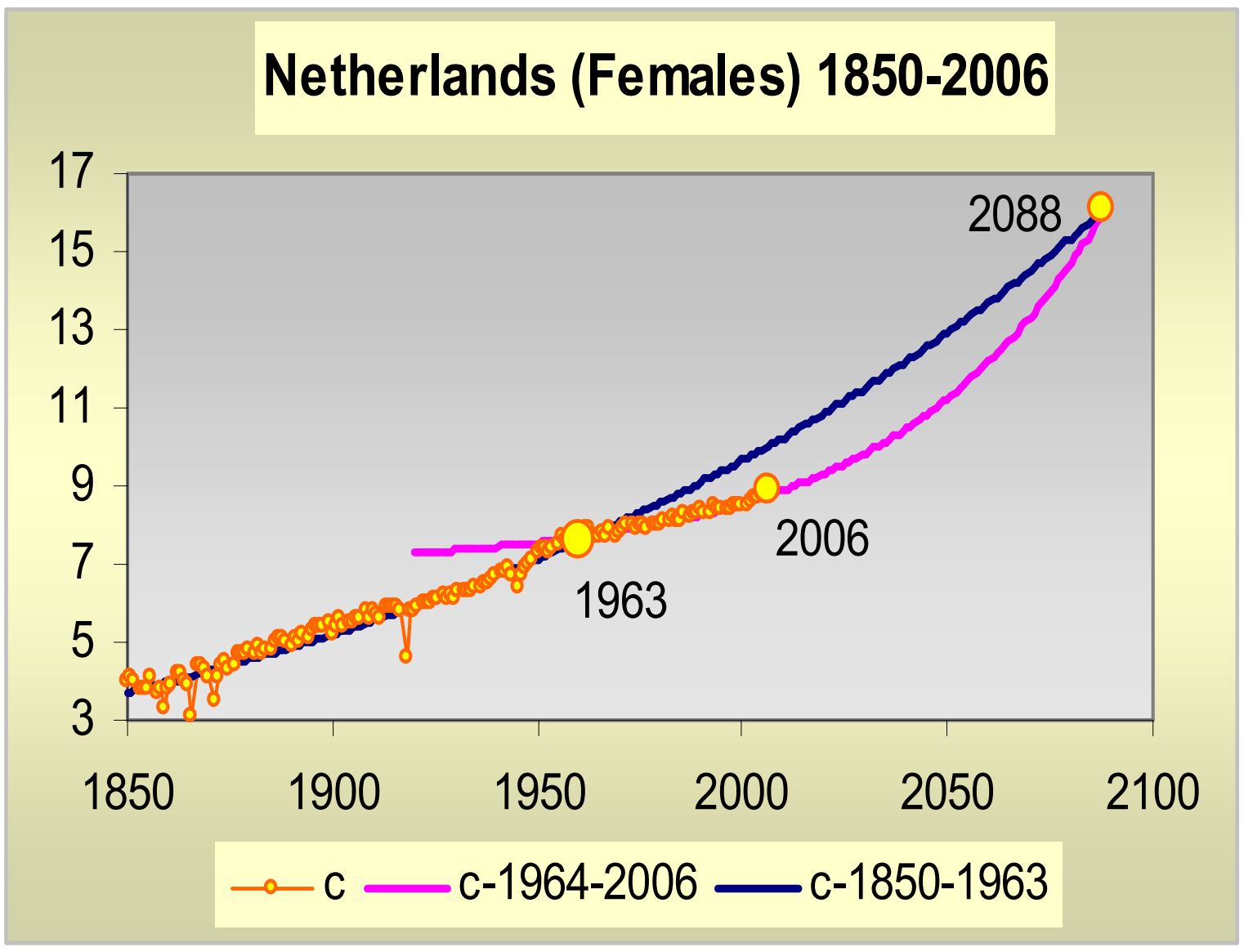

Figure 9. Exponent *c*: fit and forecasts

By estimating the future trends of the parameters *b*, *l* and *c* we can calculate the death distribution for several time periods. We have selected the parameters forecasts for the years 2020, 2040, 2060 and 2080. The values for these parameters are summarized in Table VII. The related graphs are illustrated in Figure 10.

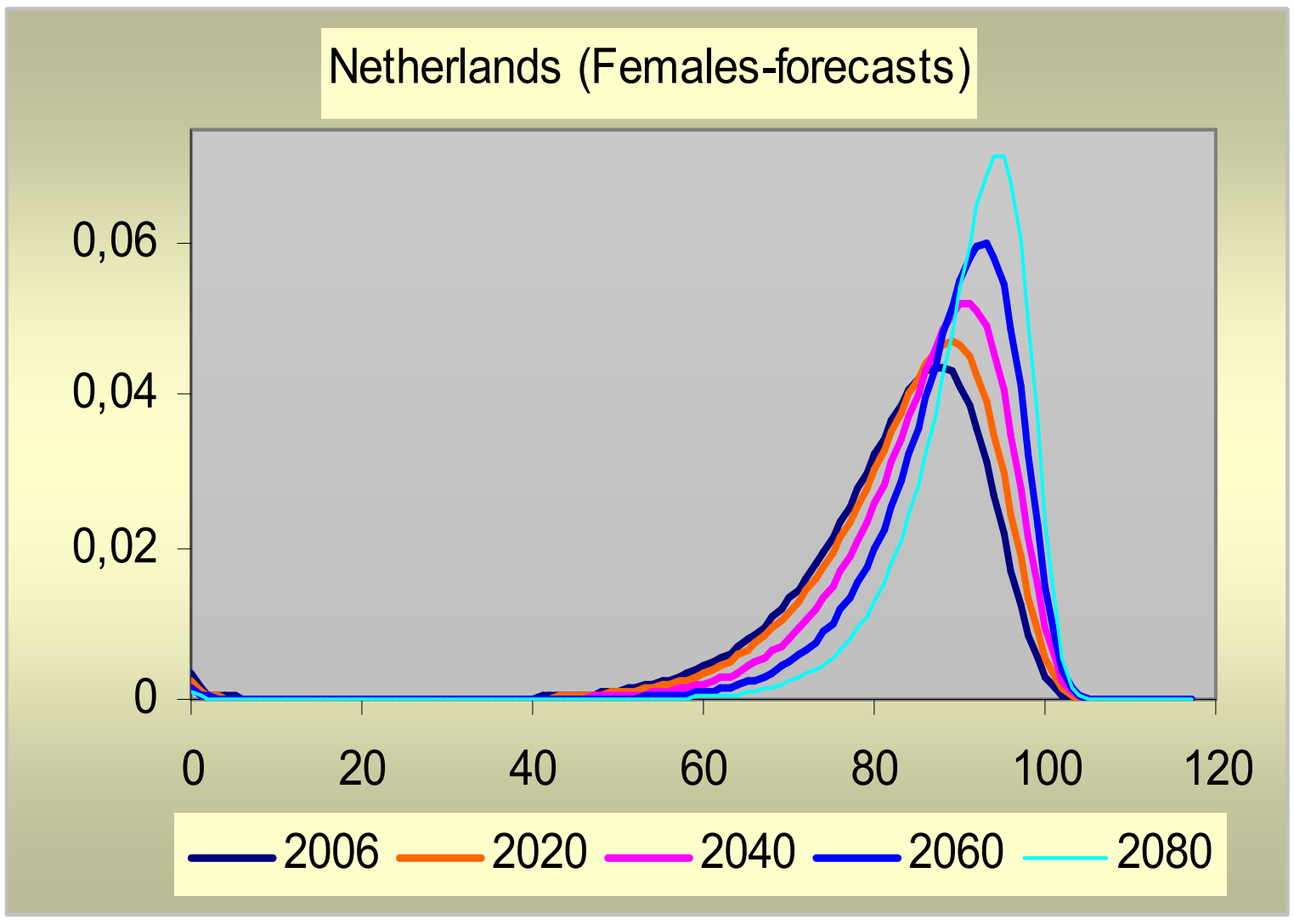

Figure 10: Forecasts of death rates in Netherlands (2020, 2040, 2060 and 2080)

TABLE VII

| Parameter | 2006 | 2020 | 2040 | 2060 | 2080 |
|---|---|---|---|---|---|
| b | 0,01443 | 0,01407 | 0,01345 | 0,01282 | 0,0122 |
| l | 0,004531 | 0,002711 | 0,00191 | 0,001574 | 0,001432 |
| c | 8,922 | 9,327 | 10,475 | 12,176 | 14,698 |

We also explore the behavior of the death distribution at the right inflection point. As it was already tested for the case of Sweden [18] the tangent at the right inflection point of the death distribution tends to shift to a position perpendicular to the X axis, whereas the right displacement of this line is slower Figure 11). In other words the characteristic tangent which we call as *the longevity tangent* [18] looks like a barrier set by the human organism to our efforts for longevity. The results for Netherlands (Females) support the previous findings for Sweden (Females).

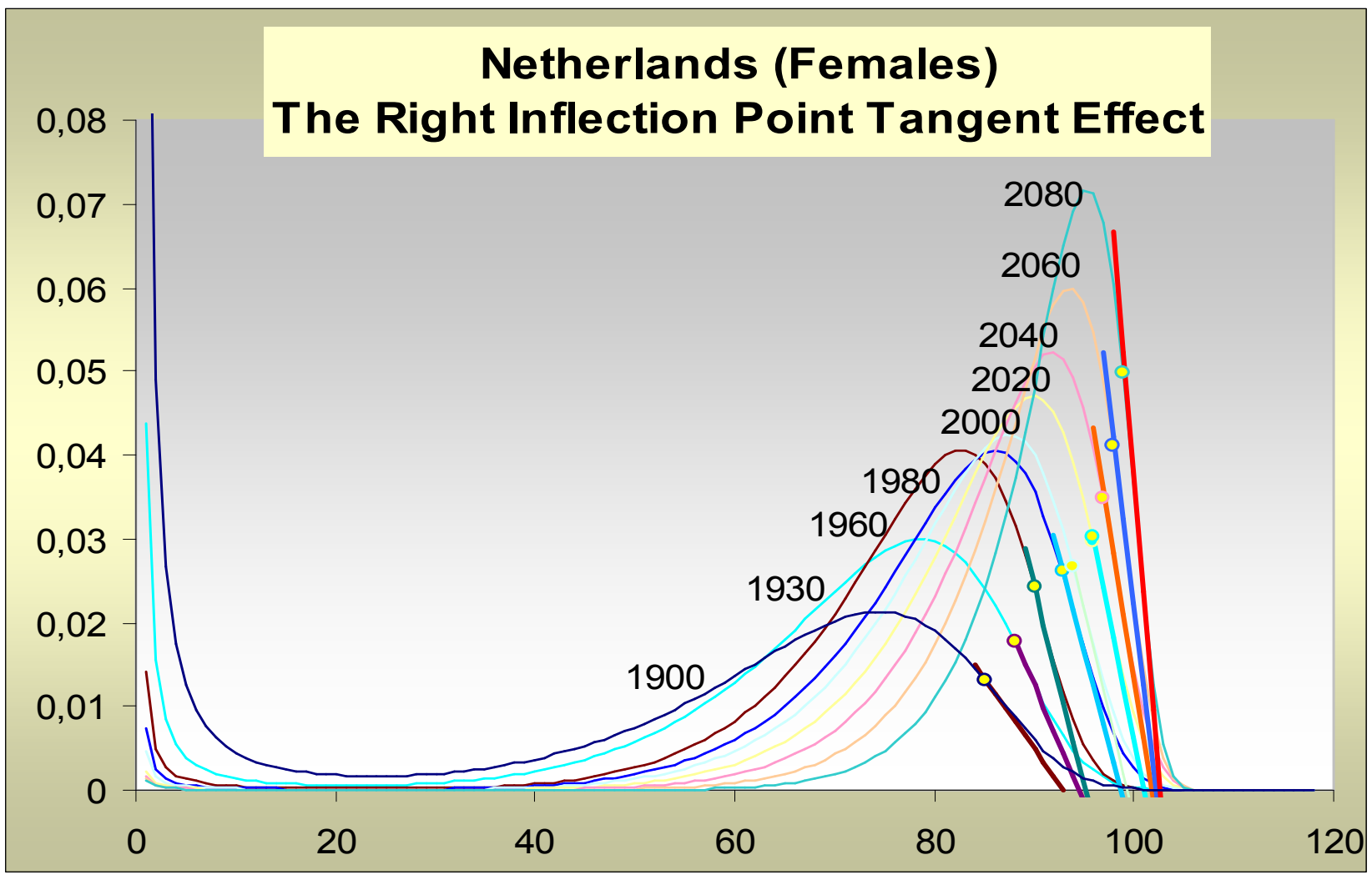

Figure 11. The vertical shift effect of the tangent at the right inflection point.

The estimates and forecasts for the parameters of the $g(t)$ provide another method for estimating the future trends for the life expectancy at birth (LEB). In Figure 12 we provide the graphs for the estimates of LEB by based on the series (`1850-2006) and (1900-2006). For both cases we fit a Logistic model to the data. This model has the form

$$LEB(t) = \frac{F}{1 + ((F - f_0)/f_0)e^{-at}}$$

The estimated parameters of the Logistic model are presented in Table VIII. For both time periods studied the growth parameter *a* obtains similar values, whereas the use of all the data points (1850-2006) gives an estimated for the life expectancy upper limit *F*=96,18 higher than the upper limit estimated from the series (1900-2006) that is *F*=88,89. The LEB, estimated by using the parameters *b*, *l* and *c*, the function $g(t)$ and the related program, gives us a graph (Figure 12) between the two other estimates. The Table IX includes the predictions based on the two methods proposed. The second method based on the analysis of the parameters gives higher future values for the life expectancy at birth. However, by combining both methods we can expect to improve forecasts and especially by making long range estimates.

TABLE VIII

| Time Period | $F$ | $f_0$ | $a$ |
|---|---|---|---|
| 1850-2006 | 96,18 | 32,53 | 0,01636 |
| 1900-2006 | 88,89 | 47,17 | 0,01676 |

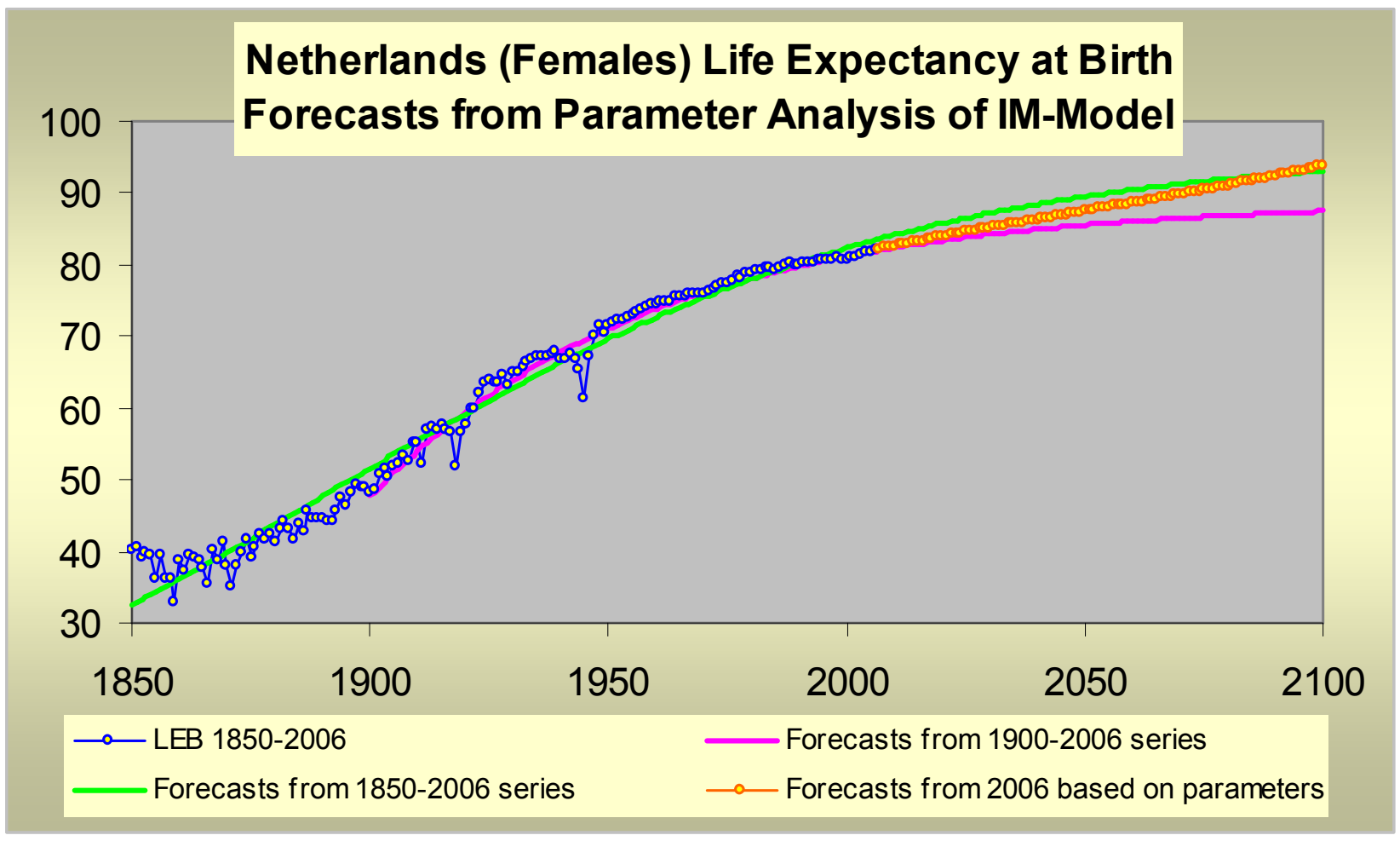

Figure 12. Estimating the Life Expectancy at Birth by using the parameter forecasts of the IM-Model (Netherlands, females)

TABLE IX

| Netherlands (Females) Life Expectancy at Birth | | | | |
|---|---|---|---|---|
| Year | Forecasts from 1900-2006 series | Forecasts from 1950-2006 series | Forecasts from 1850-2006 series | Forecasts from 2006 based on parameters |
| 2010 | 82,37 | 82,48 | 84,16 | 82,63 |
| 2015 | 82,90 | 82,95 | 85,00 | 83,28 |
| 2020 | 83,38 | 83,37 | 85,78 | 83,91 |
| 2025 | 83,82 | 83,75 | 86,51 | 84,52 |
| 2030 | 84,23 | 84,10 | 87,20 | 85,13 |
| 2035 | 84,60 | 84,42 | 87,85 | 85,72 |
| 2040 | 84,95 | 84,71 | 88,45 | 86,30 |
| 2045 | 85,27 | 84,98 | 89,01 | 86,88 |
| 2050 | 85,56 | 85,22 | 89,53 | 87,46 |
| 2055 | 85,82 | 85,44 | 90,02 | 88,04 |
| 2060 | 86,07 | 85,64 | 90,48 | 88,62 |
| 2065 | 86,30 | 85,83 | 90,90 | 89,21 |
| 2070 | 86,51 | 85,99 | 91,30 | 89,82 |
| 2075 | 86,70 | 86,14 | 91,66 | 90,43 |
| 2080 | 86,87 | 86,28 | 92,00 | 91,06 |
| 2085 | 87,04 | 86,41 | 92,32 | 91,72 |
| 2090 | 87,19 | 86,52 | 92,61 | 92,40 |
| 2095 | 87,32 | 86,63 | 92,88 | 93,10 |
| 2100 | 87,45 | 86,72 | 93,13 | 93,84 |

## Conclusions

We have applied a new theoretical framework to forecast the future life expectancy and life expectancy at birth in the Netherlands (Females). We also do predictions by based on a classical forecasting methodology. The resulting figures are close to those suggested by the Dutch Actuarial Association in the Projection Table 2010-2060 [21]. However, the two methods are based on a different underlying theory. The method used here can give reliable fitting even by estimating values in the past. By using the 1950-2006 data for Netherlands (Females) we have good predictions for the past period (1900-1950) and for the future (2006-2060) (see the Table X at the end of this paper).

## Bibliography

1. Economos, Kinetics of metazoan mortality. *J. Social Biol. Struct*., 3, 317-329 (1980).
2. Gompertz, On the nature of the function expressive of the law of human mortality, and on the mode of determining the value of life contingencies, *Philosophical Transactions of the Royal Society of London A* 115, 513-585 (1825).
3. J. Graunt, *Natural and Political Observations Made upon the Bills of Mortality*, First Edition, 1662; Fifth Edition (1676).
4. M. Greenwood and J. O. Irwin, The biostatistics of senility, *Human Biology*, vol.11, 1-23 (1939).
5. S. Haberman and T. A. Sibbett, *History of Actuarial Science*, London, UK: William Pickering, (1995).
6. E. Halley, An Estimate of the Degrees of Mortality of Mankind, Drawn from the Curious Tables of the Births and Funerals at the City of Breslau, with an Attempt to Ascertain the Price of Annuities upon Lives, *Philosophical Transactions*, Volume 17, pp. 596-610 (1693).
7. L. M. A. Heligman and J. H. Pollard, The Age Pattern of mortality, *Journal of the Institute of Actuaries* 107, part 1, 49-82 (1980).
8. J. Janssen and C. H. Skiadas, Dynamic modelling of life-table data, *Applied Stochastic Models and Data Analysis*, **11**, 1, 35-49 (1995).
9. N. Keyfitz and H. Caswell, Applied Mathematical Demography, 3rd ed., Springer (2005).
10. R. D. Lee and L. R. Carter, Modelling and forecasting U.S. mortality. J. Amer. Statist. Assoc. 87 (14), 659–675 (1992).
11. W. M. Makeham, On the Law of Mortality and the Construction of Annuity Tables, *J. Inst. Act. and Assur. Mag*. 8, 301-310, (1860).
12. H. Skiadas and C. Skiadas, A modeling approach to life table data, in *Recent Advances in Stochastic Modeling and Data Analysis*, C. H. Skiadas, Ed. (World Scientific, Singapore), 350–359 (2007).
13. C. H. Skiadas, C. Skiadas, Comparing the Gompertz Type Models with a First Passage Time Density Model, in *Advances in Data Analysis*, C. H. Skiadas Ed. (Springer/Birkhauser, Boston), 203-209 (2010).

14. C. Skiadas and C. H. Skiadas, Development, Simulation and Application of First Exit Time Densities to Life Table Data, *Communications in Statistics* 39, 444-451 (2010).
15. C. H. Skiadas and C. Skiadas, Exploring life expectancy limits: First exit time modelling, parameter analysis and forecasts, in *Chaos Theory: Modeling, Simulation and Applications*, C. H. Skiadas, I. Dimotikalis and C. Skiadas, Eds. (World Scientific, Singapore), 357–368 (2011).
16. C. H. Skiadas and C. Skiadas, The first exit time problem, in M. Lovric (editor), *International Encyclopedia of Statistical Science*, New York: Springer, 521-523 (2011), ISBN 978-3-642-04897-5.
17. C. H. Skiadas, Recent advances in stochastic modeling in M. Lovric (editor), *International Encyclopedia of Statistical Science*, New York: Springer, 1524-1526 (2011), ISBN 978-3-642-04897-5.
18. C. H. Skiadas and C. Skiadas, Properties of a stochastic model for life table data: Exploring life expectancy limits, *arxiv.org*, 9 pages, published 10-01-2011, http://arxiv.org/ftp/arxiv/papers/1101/1101.1796.pdf
19. C. H. Skiadas, A Life Expectancy Study based on the Deterioration Function and an Application to Halley's Breslau Data, *arxiv.org*, 29 pages, published 1-10-2011, http://arxiv.org/ftp/arxiv/papers/1110/1110.0130.pdf
20. B. L. Strehler and A.S. Mildvan, General theory of mortality and aging, *Science* 132, 14-21 (1960).
21. Dutch Actuarial Association, Projection Table 2010-2060, 110901.Prognosetafels.English.v1[1], (2010).

# TABLE X

| Life Expectancy at Birth (Netherlands, Females, fit and forecasts) | | | | | | | |
|---|---|---|---|---|---|---|---|
| Year | Life Exp at Birth (Data) | Life Exp at Birth (Fit) | Life Exp at Birth (Mean) | Max Det Age | Life Exp (DTR Method) | LEB Estimates, fit 1950-2006 | LEB Estimates, fit 1950-2006 |
| 1900 | 49,81 | 48,15 | 48,98 | 74,58 | 78,86 | 47,71 | 47,23 |
| 1901 | 50,10 | 48,54 | 49,32 | 75,58 | 79,46 | 48,39 | 47,98 |
| 1902 | 51,71 | 50,78 | 51,24 | 75,13 | 78,84 | 49,07 | 48,71 |
| 1903 | 52,82 | 51,38 | 52,10 | 76,40 | 80,22 | 49,73 | 49,43 |
| 1904 | 52,02 | 50,43 | 51,22 | 75,90 | 79,72 | 50,38 | 50,14 |
| 1905 | 53,28 | 51,91 | 52,60 | 75,77 | 79,59 | 51,02 | 50,83 |
| 1906 | 53,61 | 52,13 | 52,87 | 76,16 | 79,86 | 51,65 | 51,51 |
| 1907 | 54,47 | 53,39 | 53,93 | 75,42 | 79,23 | 52,27 | 52,18 |
| 1908 | 53,80 | 52,49 | 53,14 | 74,92 | 78,54 | 52,88 | 52,83 |
| 1909 | 55,86 | 55,16 | 55,51 | 75,53 | 79,26 | 53,48 | 53,48 |
| 1910 | 56,04 | 55,10 | 55,57 | 75,72 | 79,21 | 54,06 | 54,11 |
| 1911 | 54,25 | 52,27 | 53,26 | 75,73 | 79,33 | 54,64 | 54,73 |
| 1912 | 57,74 | 57,04 | 57,39 | 76,59 | 80,16 | 55,21 | 55,34 |
| 1913 | 58,06 | 57,20 | 57,63 | 76,19 | 79,59 | 55,77 | 55,93 |
| 1914 | 57,95 | 56,97 | 57,46 | 76,28 | 79,66 | 56,32 | 56,52 |
| 1915 | 58,00 | 57,85 | 57,93 | 75,63 | 79,02 | 56,86 | 57,10 |
| 1916 | 56,83 | 56,99 | 56,91 | 75,48 | 78,96 | 57,40 | 57,66 |
| 1917 | 56,40 | 56,73 | 56,57 | 75,28 | 78,84 | 57,92 | 58,21 |
| 1918 | 48,49 | 51,93 | 50,21 | 77,90 | 82,29 | 58,43 | 58,76 |
| 1919 | 55,55 | 56,58 | 56,07 | 75,39 | 78,97 | 58,94 | 59,29 |
| 1920 | 58,04 | 57,86 | 57,95 | 75,97 | 79,49 | 59,44 | 59,82 |
| 1921 | 60,20 | 59,76 | 59,98 | 76,29 | 79,62 | 59,93 | 60,33 |
| 1922 | 60,23 | 60,00 | 60,12 | 74,97 | 78,40 | 60,41 | 60,84 |
| 1923 | 62,39 | 62,22 | 62,31 | 76,64 | 79,81 | 60,88 | 61,33 |
| 1924 | 63,27 | 63,45 | 63,36 | 76,78 | 79,95 | 61,35 | 61,82 |
| 1925 | 63,61 | 64,07 | 63,84 | 76,68 | 79,80 | 61,81 | 62,29 |
| 1926 | 63,41 | 63,53 | 63,47 | 76,56 | 79,67 | 62,26 | 62,76 |
| 1927 | 63,09 | 63,41 | 63,25 | 75,60 | 78,68 | 62,70 | 63,22 |
| 1928 | 64,16 | 64,48 | 64,32 | 76,47 | 79,63 | 63,13 | 63,67 |
| 1929 | 62,63 | 63,06 | 62,85 | 75,10 | 78,23 | 63,56 | 64,12 |
| 1930 | 65,07 | 65,11 | 65,09 | 77,04 | 80,10 | 63,98 | 64,55 |
| 1931 | 64,75 | 65,02 | 64,88 | 75,67 | 78,65 | 64,40 | 64,98 |
| 1932 | 65,78 | 65,75 | 65,77 | 76,32 | 79,30 | 64,80 | 65,40 |
| 1933 | 66,47 | 66,55 | 66,51 | 76,49 | 79,42 | 65,20 | 65,81 |
| 1934 | 67,07 | 66,93 | 67,00 | 77,09 | 79,97 | 65,60 | 66,21 |
| 1935 | 67,00 | 67,08 | 67,04 | 76,45 | 79,28 | 65,99 | 66,61 |
| 1936 | 67,15 | 67,09 | 67,12 | 76,48 | 79,31 | 66,37 | 67,00 |
| 1937 | 67,49 | 67,33 | 67,41 | 76,16 | 78,93 | 66,74 | 67,38 |
| 1938 | 67,93 | 67,72 | 67,83 | 76,78 | 79,52 | 67,11 | 67,76 |
| 1939 | 68,30 | 68,01 | 68,16 | 76,44 | 79,15 | 67,47 | 68,12 |
| 1940 | 67,03 | 66,90 | 66,97 | 75,69 | 78,36 | 67,83 | 68,49 |
| 1941 | 66,66 | 66,69 | 66,68 | 75,45 | 78,01 | 68,18 | 68,84 |
| 1942 | 67,31 | 67,64 | 67,48 | 76,61 | 79,12 | 68,52 | 69,19 |
| 1943 | 65,79 | 66,82 | 66,31 | 76,44 | 78,87 | 68,86 | 69,53 |
| 1944 | 63,13 | 65,27 | 64,20 | 75,86 | 78,51 | 69,19 | 69,86 |
| 1945 | 60,85 | 61,25 | 61,05 | 75,04 | 77,96 | 69,52 | 70,19 |
| 1946 | 68,01 | 67,20 | 67,61 | 77,99 | 80,44 | 69,84 | 70,52 |
| 1947 | 70,57 | 70,07 | 70,32 | 77,69 | 80,02 | 70,16 | 70,83 |
| 1948 | 72,16 | 71,46 | 71,81 | 78,41 | 80,65 | 70,47 | 71,14 |
| 1949 | 71,43 | 70,63 | 71,03 | 76,86 | 79,08 | 70,78 | 71,45 |
| 1950 | 72,39 | 71,74 | 72,06 | 77,73 | 79,80 | 71,08 | 71,75 |
| 1951 | 72,54 | 71,86 | 72,20 | 77,79 | 79,78 | 71,37 | 72,04 |
| 1952 | 73,06 | 72,30 | 72,68 | 78,09 | 80,07 | 71,66 | 72,33 |
| 1953 | 72,77 | 72,35 | 72,56 | 78,01 | 80,04 | 71,95 | 72,62 |
| 1954 | 73,63 | 72,79 | 73,21 | 78,24 | 80,19 | 72,23 | 72,89 |
| 1955 | 73,86 | 73,11 | 73,49 | 78,30 | 80,18 | 72,51 | 73,17 |
| 1956 | 73,92 | 73,24 | 73,58 | 77,88 | 79,67 | 72,78 | 73,44 |
| 1957 | 74,39 | 73,79 | 74,09 | 78,51 | 80,33 | 73,05 | 73,70 |
| 1958 | 74,63 | 73,96 | 74,30 | 78,61 | 80,39 | 73,31 | 73,96 |
| 1959 | 74,94 | 74,39 | 74,67 | 78,83 | 80,52 | 73,57 | 74,21 |
| 1960 | 75,13 | 74,47 | 74,80 | 78,84 | 80,56 | 73,83 | 74,46 |
| 1961 | 75,51 | 74,78 | 75,15 | 79,15 | 80,85 | 74,08 | 74,70 |
| 1962 | 75,38 | 74,83 | 75,10 | 78,97 | 80,62 | 74,32 | 74,94 |
| 1963 | 75,59 | 75,02 | 75,30 | 79,09 | 80,76 | 74,56 | 75,18 |
| 1964 | 76,07 | 75,54 | 75,81 | 79,79 | 81,50 | 74,80 | 75,41 |
| 1965 | 75,99 | 75,46 | 75,72 | 79,56 | 81,27 | 75,04 | 75,64 |
| 1966 | 75,96 | 75,58 | 75,77 | 79,58 | 81,25 | 75,27 | 75,86 |
| 1967 | 76,40 | 75,97 | 76,19 | 80,06 | 81,74 | 75,49 | 76,08 |
| 1968 | 76,28 | 76,01 | 76,15 | 79,77 | 81,38 | 75,72 | 76,29 |
| 1969 | 76,10 | 75,78 | 75,94 | 79,85 | 81,55 | 75,93 | 76,50 |

TABLE X (continued)

| | | | | | | | |
|---|---|---|---|---|---|---|---|
| 1970 | 76,37 | 76,12 | 76,24 | 79,96 | 81,62 | 76,15 | 76,71 |
| 1971 | 76,66 | 76,41 | 76,54 | 79,94 | 81,55 | 76,36 | 76,91 |
| 1972 | 76,69 | 76,50 | 76,60 | 79,89 | 81,47 | 76,57 | 77,11 |
| 1973 | 77,03 | 77,04 | 77,04 | 80,48 | 82,03 | 76,77 | 77,31 |
| 1974 | 77,50 | 77,31 | 77,40 | 80,84 | 82,38 | 76,98 | 77,50 |
| 1975 | 77,57 | 77,42 | 77,50 | 80,71 | 82,21 | 77,17 | 77,69 |
| 1976 | 77,75 | 77,58 | 77,66 | 80,98 | 82,46 | 77,37 | 77,87 |
| 1977 | 78,32 | 78,31 | 78,31 | 81,74 | 83,24 | 77,56 | 78,05 |
| 1978 | 78,29 | 78,30 | 78,29 | 81,55 | 83,00 | 77,75 | 78,23 |
| 1979 | 78,71 | 78,73 | 78,72 | 82,00 | 83,44 | 77,93 | 78,40 |
| 1980 | 78,95 | 78,80 | 78,87 | 82,12 | 83,55 | 78,12 | 78,58 |
| 1981 | 79,11 | 79,07 | 79,09 | 82,22 | 83,61 | 78,29 | 78,74 |
| 1982 | 79,24 | 79,23 | 79,24 | 82,27 | 83,63 | 78,47 | 78,91 |
| 1983 | 79,40 | 79,45 | 79,43 | 82,51 | 83,84 | 78,64 | 79,07 |
| 1984 | 79,50 | 79,42 | 79,46 | 82,55 | 83,91 | 78,81 | 79,23 |
| 1985 | 79,49 | 79,36 | 79,43 | 82,47 | 83,83 | 78,98 | 79,39 |
| 1986 | 79,44 | 79,50 | 79,47 | 82,39 | 83,69 | 79,15 | 79,54 |
| 1987 | 79,90 | 79,99 | 79,94 | 82,97 | 84,27 | 79,31 | 79,69 |
| 1988 | 80,09 | 80,16 | 80,12 | 82,86 | 84,15 | 79,47 | 79,84 |
| 1989 | 79,78 | 80,03 | 79,90 | 82,73 | 84,01 | 79,62 | 79,98 |
| 1990 | 79,95 | 80,09 | 80,02 | 82,77 | 84,02 | 79,78 | 80,13 |
| 1991 | 80,01 | 80,26 | 80,13 | 82,92 | 84,19 | 79,93 | 80,27 |
| 1992 | 80,15 | 80,39 | 80,27 | 83,07 | 84,33 | 80,08 | 80,40 |
| 1993 | 79,87 | 80,24 | 80,06 | 82,63 | 83,86 | 80,23 | 80,54 |
| 1994 | 80,18 | 80,52 | 80,35 | 82,98 | 84,21 | 80,37 | 80,67 |
| 1995 | 80,24 | 80,63 | 80,43 | 82,98 | 84,21 | 80,51 | 80,80 |
| 1996 | 80,22 | 80,52 | 80,37 | 83,00 | 84,24 | 80,65 | 80,93 |
| 1997 | 80,41 | 80,79 | 80,60 | 83,14 | 84,34 | 80,79 | 81,05 |
| 1998 | 80,55 | 80,89 | 80,72 | 83,14 | 84,33 | 80,92 | 81,17 |
| 1999 | 80,32 | 80,70 | 80,51 | 83,05 | 84,25 | 81,05 | 81,30 |
| 2000 | 80,44 | 80,79 | 80,62 | 83,05 | 84,23 | 81,18 | 81,41 |
| 2001 | 80,59 | 80,92 | 80,76 | 83,17 | 84,35 | 81,31 | 81,53 |
| 2002 | 80,57 | 80,99 | 80,78 | 83,13 | 84,29 | 81,44 | 81,64 |
| 2003 | 80,81 | 81,32 | 81,07 | 83,25 | 84,36 | 81,56 | 81,75 |
| 2004 | 81,32 | 81,72 | 81,52 | 83,64 | 84,74 | 81,68 | 81,86 |
| 2005 | 81,48 | 81,85 | 81,66 | 83,78 | 84,84 | 81,80 | 81,97 |
| 2006 | 81,77 | 82,15 | 81,96 | 83,88 | 84,90 | 81,92 | 82,08 |
| 2007 | | | | | | 82,04 | 82,18 |
| 2008 | | | | | | 82,15 | 82,28 |
| 2009 | | | | | | 82,26 | 82,38 |
| 2010 | | | | | | 82,37 | 82,48 |
| 2011 | | | | | | 82,48 | 82,58 |
| 2012 | | | | | | 82,59 | 82,67 |
| 2013 | | | | | | 82,69 | 82,76 |
| 2014 | | | | | | 82,80 | 82,86 |
| 2015 | | | | | | 82,90 | 82,95 |
| 2016 | | | | | | 83,00 | 83,03 |
| 2017 | | | | | | 83,09 | 83,12 |
| 2018 | | | | | | 83,19 | 83,20 |
| 2019 | | | | | | 83,29 | 83,29 |
| 2020 | | | | | | 83,38 | 83,37 |
| 2021 | | | | | | 83,47 | 83,45 |
| 2022 | | | | | | 83,56 | 83,53 |
| 2023 | | | | | | 83,65 | 83,60 |
| 2024 | | | | | | 83,74 | 83,68 |
| 2025 | | | | | | 83,82 | 83,75 |
| 2026 | | | | | | 83,91 | 83,83 |
| 2027 | | | | | | 83,99 | 83,90 |
| 2028 | | | | | | 84,07 | 83,97 |
| 2029 | | | | | | 84,15 | 84,04 |
| 2030 | | | | | | 84,23 | 84,10 |
| 2031 | | | | | | 84,31 | 84,17 |
| 2032 | | | | | | 84,38 | 84,24 |
| 2033 | | | | | | 84,46 | 84,30 |
| 2034 | | | | | | 84,53 | 84,36 |
| 2035 | | | | | | 84,60 | 84,42 |
| 2036 | | | | | | 84,68 | 84,48 |
| 2037 | | | | | | 84,75 | 84,54 |
| 2038 | | | | | | 84,81 | 84,60 |
| 2039 | | | | | | 84,88 | 84,66 |
| 2040 | | | | | | 84,95 | 84,71 |
| 2041 | | | | | | 85,01 | 84,77 |
| 2042 | | | | | | 85,08 | 84,82 |
| 2043 | | | | | | 85,14 | 84,88 |
| 2044 | | | | | | 85,20 | 84,93 |